\documentclass{article}
\usepackage{spconf,amsmath,graphicx}

\ninept
\title{Real-Time MRI Video synthesis from time aligned phonemes with sequence-to-sequence networks}

\name{Sathvik Udupa, Prasanta Kumar Ghosh}
\address{Electrical Engineering Department, Indian Institute of Science (IISc), Bangalore - 560012, India.}

\begin{document}
\maketitle
\begin{abstract}
Real-Time Magnetic resonance imaging (rtMRI) of the midsagittal plane of the mouth is of interest for speech production research. In this work, we focus on estimating utterance level rtMRI video from the spoken phoneme sequence. We obtain time-aligned phonemes from forced alignment, to obtain frame-level phoneme sequences which are aligned with rtMRI frames. We propose a sequence-to-sequence learning model with a transformer phoneme encoder and convolutional frame decoder. We then modify the learning by using intermediary features obtained from sampling from a pretrained phoneme-conditioned variational autoencoder (CVAE). We train on 8 subjects in a subject-specific manner and demonstrate the performance with a subjective test. We also use an auxiliary task of air tissue boundary (ATB) segmentation to obtain the objective scores on the proposed models. We show that the proposed method is able to generate realistic rtMRI video for unseen utterances, and adding CVAE is beneficial for learning the sequence-to-sequence mapping for subjects where the mapping is hard to learn.
\end{abstract}


\section{Introduction}
\label{sec:intro}
In recent years, real-time Magnetic Resonance Imaging (rtMRI) has been emerging as an important tool in understanding speech production \cite{rtmri_recent}. The rtMRI of a speaker allows visualising the dynamics of various speech articulators, irrespective of language or health conditions. While other data sources such as Electromagnetic articulography (EMA) \cite{ema_old}, X-ray \cite{xray} and Ultrasound \cite{ultrasound} can provide data at a higher frequency, rtMRI captures various regions in the midsaggital plane such as the pharynx, larynx, velum along with jaw and the articulators in the oral cavity. EMA tracks sensors on a few articulators in the mouth, hence a complete view of different articulators is not present. Similarly, ultrasound imaging doesn't capture structures such as palate or pharyngeal walls \cite{preston2017ultrasound}.
rtMRI is also non-invasive, which is not the case for EMA. Additionally, there has been work \cite{lingala2017fast} at extracting rtMRI frames at a higher sample rate. 

The rtMRI allows for understanding the movement and coordination of various speech articulators. Visualisation of articulators been used to aid in learning pronunciation \cite{pronunciation, toutios2016illustrating} and air tissue boundaries obtained from rtMRI has been used in various applications such as speech synthesis \cite{tts_with_rtmri}, speaker verification \cite{spk_verification}, phone classification \cite{ph_classify_from_rtmri} etc. Different works \cite{rtmri_articulation_example} have also looked at understanding articulation using rtMRI data.

\begin{figure*}[h]
    \centering
    \includegraphics[width=18cm]{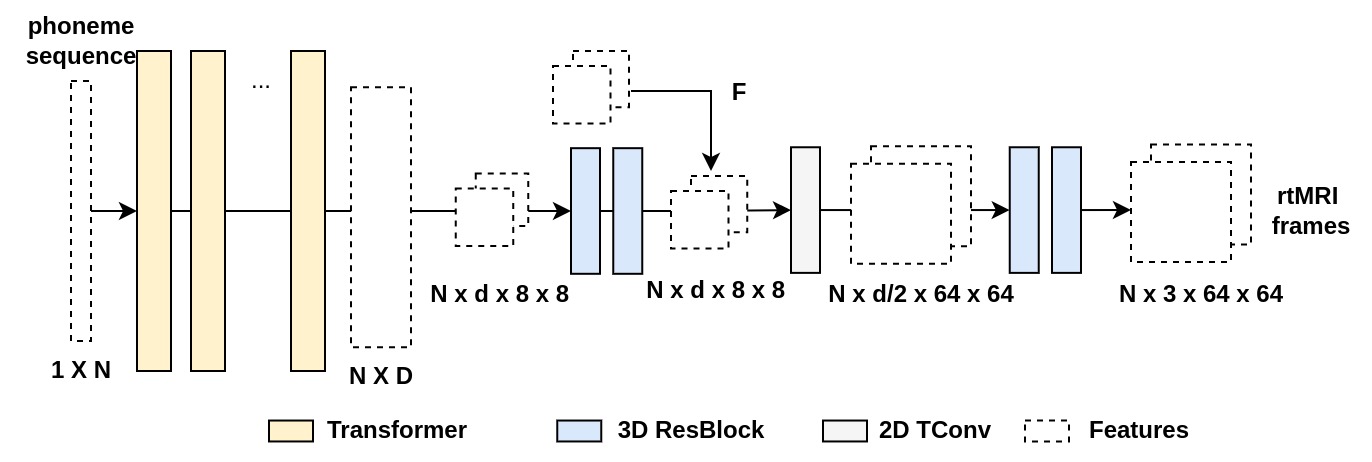}
    \caption{Figure represents the model used for synthesizing rtMRI video from phonemes. The transformer acts as an encoder which represents phonemes in a \emph{D} dimensional space. These features are then reshaped as 
 \emph{d} feature maps of size \emph{8} x \emph{8}, where \emph{D} = \emph{d} * \emph{8} * \emph{8}. It is then passed through a series of 3D (ResBlock) and 2D (TConv) convolutional layers in order to produce the frames. Dotted shapes represent the features after different layers. \emph{F} represents the features from the decoder from CVAE in \emph{s2s-v} model.}
    \label{fig:model}
\end{figure*}
While rtMRI has its benefits, there is still considerable effort and apparatus required to collect rtMRI data. Hence, there have been efforts to synthesise rtMRI frames from other modalities. Previously, there have been works on obtaining word level rtMRI frames using concatenative approaches\cite{previous_text2rtmri}, \cite{previous_text2rtmri_withatb}. In these methods, a phone-specific image repository is created from the training set by aligning frames to phonemes based on forced alignment. During test time, frames are selected for the required phonemes by employing a dynamic programming approach to ensure smoothness across frames. This is followed by interpolation to match the duration as required by forced alignment. The synthesis is performed on a word level and the authors found that the synthesis procedure was able to produce a realistic video but having more phonemes in a word reduces the quality of synthesis. Hence, it is not ideal for longer words or full utterances. 

We explore the use of end-to-end deep learning models in order to learn the mapping from phonemes to rtMRI video, on a full sequence. Deep learning methods are appropriate for this task since the models can learn across long sequences of arbitrarily different lengths, and can learn between modalities and generalise well to unseen situations. In recent years, transformer neural networks have been shown to perform well on various speech-based tasks involving sequence-to-sequence modelling such as speech synthesis \cite{fastspeech}, acoustic to articulatory inversion \cite{udupa_fastspeech} etc. Self-attention operation in transformer networks enables learning the dependencies between positions in a sequence, making it suitable for learning good phoneme-level features. Convolutional neural networks have been commonly used as image decoders in various image prediction tasks such as U-Net\cite{unet}, SegNet\cite{badrinarayanan2017segnet} and combinations of 2D and 3D convolution have been utilised to learn spatial and temporal information in video prediction tasks \cite{3d2d_video2video}. A mix of 2D and 3D convolutions is suitable for decoder since rtMRI synthesis requires the articulators to be anatomically accurate w.r.t. the uttered phoneme while having smooth transitions in phoneme boundaries.  We also incorporate a conditional variational autoencoder (CVAE) to incorporate frame-level features which encode prior information on articulators for corresponding phonemes. This type of sample-conditioned training methodology has helped in image prediction in specific domains requiring anatomical correctness such as cardiac image segmentation \cite{cvae_medical}.

To summarise, we treat phoneme to rtMRI synthesis as a sequence-to-sequence prediction task. We use a transformer neural network as an encoder to represent the phonemes of the full input sequence and use a series of convolutional neural networks as decoders to predict the rtMRI frames. We also add to this approach by conditioning the image decoder with intermediary features from a pretrained CVAE, to aid in learning the mapping. We perform a subjective evaluation on the synthesis of all subjects' data, along with an objective evaluation with dice score on air tissue boundary (ATB) masks from a pretrained SegNet, which is trained on rtMRI to ATB masks.

\section{Dataset}
We use rtMRI data from USC-TIMIT \cite{usctimit} speech production database in our work. The dataset consists of rtMRI data of 10 American English speakers (5 male and 5 female). The MRI is obtained at a frame rate of 23.18 with frames of shape \emph{68} x \emph{68} x \emph{3}. The audio is recorded simultaneously at 20kHz frequency, from 460 utterances from the MOCHA-TIMIT corpus. \cite{mochatimit}. Phoneme alignment is obtained by performing forced alignment with Kaldi \cite{povey2011kaldi}, and the aligned is processed to align with rtMRI frames at 23.18 Hz. In our experiments, we use data from 8 speakers - M1, M2, M3, M4, F1, F2, F3, F4. Two speakers were left out for unseen speaker studies.
\label{sec:format}

\section{Proposed Methodology}
\label{sec:pagestyle}
We consider the problem of rtMRI synthesis as a sequence-to-sequence problem with input text sequence of length \emph{n} and output rtmRI video of shape \emph{n} x \emph{3} x \emph{64} x \emph{64}. As shown in Fig. \ref{fig:model}, we use a transformer architecture to represent the input phoneme sequence. Transformers are an ideal choice since they have been shown to perform well in learning text features, especially in multi-modal scenarios \cite{fastspeech, udupa_fastspeech}. We then use a custom convolutional decoder consisting of 2D and 3D convolutional neural networks (CNN) to represent the frames. The CNN decoder is designed to upsample the representation to the image size while maintaining the temporal context learnt from the transformer encoder. In the following sections, we go into detail about the encoder and decoder architecture.

\subsubsection{Transformer Encoder:}
We use 12 transformer layers as the encoder for the sequence-to-sequence model. The input phoneme labels are vectorised with a learnable embedding which is then given to the transformer. Transformer layers start with a self-attention layer. In self-attention, three learnable projections of the input called key, query and value are obtained. The dot product is taken between the key and query, followed by softmax activation. This acts as the attention matrix which learns dependencies between phonemes. This is then multiplied by the value matrix, which results in learning the representation of the utterance. After self-attention, the results are passed through feed-forward networks to combine the information learnt in the self-attention operation. There are additional features such as residual connection, layer normalisation and positional encoding which make the transformer a powerful representation learner. The final transformer layer produces the features with \emph{D} dimension, which is used to extract frame-level features in the decoder.

\subsubsection{Convolutional Decoder:}
With the output of transformer encoder, we first rearrange the features as \emph{n} x \emph{d} x \emph{8} x \emph{8} which can be considered as feature maps representing low level frame features. This is then passed through two 3D convolution layers with residual connections (ResBlock) to represent the temporal information learnt from the transformer after being reshaped as an image. Each ResBlock contains two sets of 3D convolution layers with a kernel size of 3, batch normalisation and ReLU activation. 3D CNN learns over the sequence, with which it learns to retain temporal information. The ResBlock output is then upsampled with two transposed convolution layers, which is represented as Tconv in Fig. \ref{fig:model}. Finally, two more ResBlocks are used to maintain the temporal information at the output and reduce the number of channels to 3 as present in the image. These series of 2D and 3D convolutional layers act as our decoder capable of producing realistic rtMRI frames while retaining temporal context from the input phoneme sequence. We label this sequence to sequence model as \emph{s2s}.

\subsubsection{Variational prior:}
\begin{figure}[h]
    \centering
    \includegraphics[width=8cm]{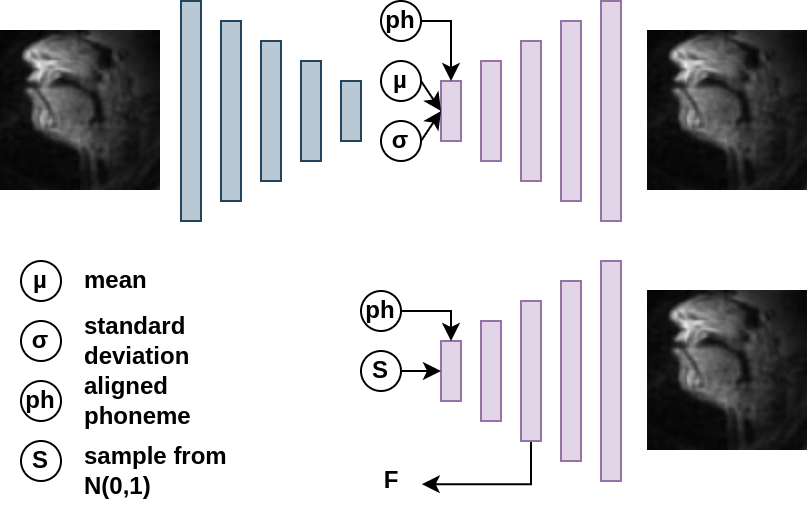}
    \caption{Figure shows the use of phoneme conditional variational autoencoder. Training is done by conditioning latent space with phoneme labels. During inference, given phoneme and sample from the normal distribution, a frame is synthesised. An intermediary feature \emph{F} is used for \emph{s2s-v}.}
    \label{fig:cvae}
\end{figure}

 In our experiments, we found that the \emph{s2s} model works well in most cases but fails to produce realistic frames in subject M1. In order to avoid such issues and provide context for the model to learn the multi-modal mapping, we utilise features from conditional variational autoencoder (CVAE) \cite{vae} as shown in Fig. \ref{fig:cvae}. CVAE is used as a lightweight convolution-based generative model, trained on frames of rtMRI, with the VAE latent space conditioned on the phoneme corresponding to the frame. It is optimised with the ELBO loss as well as reconstruction loss on frames \cite{vae}. This allows us to conditionally sample frames from phoneme labels. For all the phonemes in the \emph{s2s} model, we extract the representation from the 3nd decoder layer (denoted as \emph{F} in Fig. \ref{fig:cvae}) of CVAE in the sampling process, and these features are concatenated after the first 3D CNN block in the decoder of \emph{s2s}, as shown in Fig. \ref{fig:model}. We label this variant of \emph{s2s} as \emph{s2s-v}.  
 
It is well known that VAEs tend to produce blurry images. 
Also, while the predicted images from CVAE might represent the shape expected for a phoneme, it will not be able to predict the rtMRI frame with context from neighbouring phonemes. Due to these reasons, we make a design choice to extract an intermediary representation in the sampling process from CVAE, rather than the output image. These extracted features denoted by \emph{F}, are concatenated in \emph{s2s-v} as shown in Fig. \ref{fig:model}.

\section{Experimental Setup}
\label{sec:typestyle}
We use data from 8 subjects (4 male, 4 female) and train subject-specific models. We use 80\% of data for training, 10\% for validation and the remaining 10\% for testing. The sentences used for all the splits are common across subjects. Hence, we train 8 models and evaluate on the same speaker, unseen sentence configuration. We resize all images to \emph{64} x \emph{64} to simplify the upsampling steps in decoder and normalise them between [0,1]. The input phoneme sequence is aligned with rtMRI frames such that there is a one-to-one mapping between the two modalities.

For training CVAE, we consider the data at the frame level and shuffle all the frames within the training and validation set. The phoneme aligned to each frame is used as the conditioning in the latent space of VAE as shown in Fig. \ref{fig:cvae}. We use variational and reconstruction loss to train the VAE and train it only for 5 epochs since it converges at a very fast pace and is able to generate good frames within this time period.

For training the sequence-to-sequence model, we train on full videos, with a batch size of 1. This is performed since the memory usage of a single video can be significant and training one sequence at a time avoids zero padding. The transformer encoder and the 3D ResBlock learn as the batch size of 1, whereas, the 2D transpose convolution (TConv) responsible for upsampling, views the sequence length as a batch, then learns and upsamples over independent frames. The final 3D ResBlock again acts on a batch size of 1, applying 3D convolution on the sequence. We use a Mean Squared Error loss for \emph{s2s} and \emph{s2s-v} models. 
All models are trained with PyTorch, and the Adam optimizer is used with a learning rate of 0.0001. The checkpoint is stored based on early stopping criteria, i.e. when validation loss does not decrease after a few epochs.
For evaluating the synthesized video, we use objective as well as subjective metrics. 
\begin{figure}
    \centering
    \includegraphics[width=9cm]{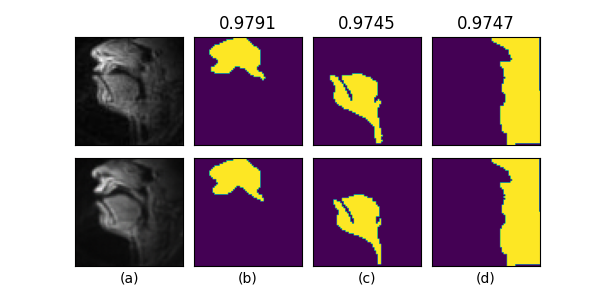}
    \vspace{-0.6cm}
    \caption{Figure represents the 3 masks used for ATB segmentation from rtMRI frames, the first row corresponds to a frame from a real rtMRI video, and the second row corresponds to the predicted frame from \emph{s2s} model for the same utterance in subject F4. (a) mask1 - predicted mask from both frames for the region above the palate. (b) mask2 - predicted mask from both frames which capture the tongue, jaw etc. (c) mask3 - predicted masks from both frames in the region of the pharyngeal wall. dice scores between the frames are shown at the top.}
    \label{fig:segments}
\end{figure}
\textbf{Objective evaluation:}

We train a SegNet \cite{badrinarayanan2017segnet, valliappan2019improved} from ground truth frames to air tissue boundary (ATB) masks. We use the SegNet variation from \cite{anwesha_dice} which has one encoder and three decoders, which predicts three ATB masks in a single model. The SegNet model is trained on individual rtMRI frames, from nine rtMRI videos from each subject after resizing the data to \emph{64} x \emph{64} to match the \emph{s2s} shapes. It is optimised with cross-entropy loss over the mask labels. We first train a pooled model combining data from 10 subjects, then fine-tune it on the frame-mask pair of each subject. Using this fined-tuned subject-specific model, we predict the masks from real rtMRI video and \emph{s2s}, \emph{s2s-v} model predictions. We then compute the dice score between the masks from real videos and the masks from the proposed method. Dice loss has been previously used to improve the segmentation of ATB masks \cite{anwesha_dice}. Dice score is a similarity coefficient which quantifies spatial overlap and provides a score between [0,1], with 1 being an exact match. This objective is an indicator of the accuracy of positions and shapes of various articulators in the synthesised videos. Fig. \ref{fig:segments} represents the three different ATB masks on which the score is reported. We compute the dice score across all the frames in an utterance, and then take the average across all the utterances in the test set of a subject. With this setup, we report the objective evaluation scores for all subjects, for 3 ATB masks shown in Fig. \ref{fig:segments}.

\textbf{Subjective evaluation:}

With subjective evaluation, we aim to measure the capability of the models to produce realistic rtMRI video on the full sequence, when compared to the actual videos. We also look at identifying the need for \emph{s2s-v} variation of the \emph{s2s} model. Hence, for the subjective test, we show the ground truth reference for an utterance and as well as the videos produced by \emph{s2s} and \emph{s2s-v} for the same utterance. The evaluator has the select the video which resembles closely to the ground truth. Along with it, the evaluator also rates the selected video on a scale of 1 to 5, with 1 being of poor quality and 5 reflecting excellent quality. The test was conducted on an interface built with Jupyter notebook in python. The evaluator could play any video, any number of times. They were encouraged to focus on the movements of various articulators and score accordingly. We performed this test with 10 evaluators, each of them scoring on the synthesis of 5 random sentences from the test set for each speaker. GitHub codes for experiments and generated samples are available at https://github.com/bloodraven66/text\_to\_rtmri\_synthesis.

\section{Results and Discussion}
\label{sec:majhead}

\subsection{Subjective evaluation}

As shown in the table, we conduct two types of subjective tests. Given the reference video identified as a reference, and provided with samples from the 2 models, the evaluator has to choose the sample which resembles the reference closely. Further, the evaluator rates the chosen sample. From Table \ref{mos_table}, we can observe that subject M1 \emph{s2s-v} is heavily favoured. For other subjects, we find a fairly consistent preference for both models. Looking at the mean opinion score, we observe that M1 has a lower score of 3.39 compared to the rest of the speakers, and this score was possible with the \emph{s2s-v} model. Apart from this, we have MOS score greater than 3.67 for each of the 7 other speakers. With these numbers, we can infer that both models can be fairly consistent and produce realistic videos. Additionally, \emph{s2s-v} models are beneficial when \emph{s2s} models are not able to produce realistic frames.

\begin{table}
\centering
\caption{The first two rows contain the ABX testing preference between the 2 models. For example, for subject M1, \emph{s2s-v} was preferred over \emph{s2s} in 48 evaluations. The 3rd row contains the mean opinion score (MOS) over the preferred model.}
\label{mos_table}
\setlength{\tabcolsep}{0.15em} 
\begin{tabular}{|c|c|c|c|c|c|c|c|c|} 
\hline
sub     & M1                                                    & M2                                                    & M3                                                  & M4                                                    & F1                                                    & F2                                                   & F3                                                   & F4                                                    \\ 
\hline
\#s2s   & 2                                                     & 22                                                    & 28                                                  & 23                                                    & 30                                                    & 25                                                   & 20                                                   & 21                                                    \\
\#s2s-v & 48                                                    & 28                                                    & 22                                                  & 27                                                    & 20                                                    & 25                                                   & 30                                                   & 29                                                    \\ 
\hline
MOS     & \begin{tabular}[c]{@{}c@{}}3.396\\(0.91)\end{tabular} & \begin{tabular}[c]{@{}c@{}}3.929\\(0.85)\end{tabular} & \begin{tabular}[c]{@{}c@{}}4.0\\(0.85)\end{tabular} & \begin{tabular}[c]{@{}c@{}}3.741\\(0.89)\end{tabular} & \begin{tabular}[c]{@{}c@{}}3.833\\(0.82)\end{tabular} & \begin{tabular}[c]{@{}c@{}}4.16\\(0.61)\end{tabular} & \begin{tabular}[c]{@{}c@{}}3.67\\(0.83)\end{tabular} & \begin{tabular}[c]{@{}c@{}}3.75\\(0.77)\end{tabular}  \\
\hline
\end{tabular}
\vspace{-0.5cm}
\end{table}

\subsection{Objective evaluation}

Table \ref{objective_eval} represents the dice score evaluation on the 3 masks for both \emph{s2s} and \emph{s2s-v} models compared against the ATB predictions from the real rtMRI samples. We can observe that for subject M1, there is a slight increase in the score in all cases, with \emph{s2s-v} dice scores being higher. The difference in the values in M1 between the two models is higher compared to other subjects. This suggests the preference given to M1 in subjective scoring. Overall, all values are greater than 0.89, suggesting good overlap between the masks of ground truth videos and predicted rtMRI frames which indicate that the synthesised frames have realistic shapes of different articulators. 

\begin{figure}[h]
    \centering
    \includegraphics[width=9cm, height=4.5cm]{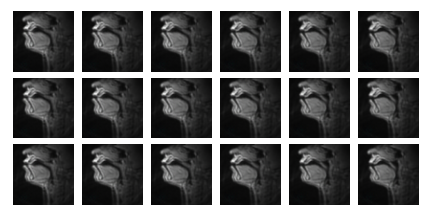}
    \caption{The figure contains consecutive samples from 1 utterance of subject F4, showing the frames from phonemes. Row 1 represents the reference video, row 2 is the seq2seq model prediction, row 3 is seq2seq-v prediction. We can observe the lowering of velum due to nasal sound being produced.}
    \label{fig:predicts}
\end{figure}

Fig. \ref{fig:predicts} shows the reference as well as predicted rtMRI frames from both the \emph{s2s} and \emph{s2s-v} models. We can observe the movement of specific articulators such as velum based on the spoken phoneme. We can observe both the \emph{s2s} and \emph{s2s-v} model results are extremely similar, indicating \emph{s2s} models are good enough in most cases and \emph{s2s-v} is useful when \emph{s2s} fails.




\begin{table}
\centering
\caption{SegNet mask DICE score on reference rtMRI frames and models, averaged across test set utterances (standard deviation in brackets)}
\label{objective_eval}
\setlength{\tabcolsep}{0.36em} 
\begin{tabular}{|c|ccc|ccc|} 
\hline
                          & \multicolumn{3}{c|}{seq2seq}                                                                                                                                                 & \multicolumn{3}{c|}{seq2seq + vae}                                                                                                                                           \\ 
\hline
\multicolumn{1}{|l|}{sub} & mask1                                                    & mask2                                                   & mask3                                                   & mask1                                                   & mask2                                                   & mask3                                                    \\ 
\hline
M1                        & \begin{tabular}[c]{@{}c@{}}0.9492\\(0.011)\end{tabular}  & \begin{tabular}[c]{@{}c@{}}0.9176\\(0.011)\end{tabular} & \begin{tabular}[c]{@{}c@{}}0.9889\\(0.001)\end{tabular} & \begin{tabular}[c]{@{}c@{}}0.9522\\(0.010)\end{tabular} & \begin{tabular}[c]{@{}c@{}}0.9237\\(0.013)\end{tabular} & \begin{tabular}[c]{@{}c@{}}0.9892\\(0.001)\end{tabular}  \\ 
\hline
M2                        & \begin{tabular}[c]{@{}c@{}}0.9667\\(0.007)\end{tabular}  & \begin{tabular}[c]{@{}c@{}}0.9352\\(0.013)\end{tabular} & \begin{tabular}[c]{@{}c@{}}0.9924\\(0.001)\end{tabular} & \begin{tabular}[c]{@{}c@{}}0.9673\\(0.007)\end{tabular} & \begin{tabular}[c]{@{}c@{}}0.9368\\(0.012)\end{tabular} & \begin{tabular}[c]{@{}c@{}}0.9922\\(0.001)\end{tabular}  \\ 
\hline
M3                        & \begin{tabular}[c]{@{}c@{}}0.9582\\(0.008)\end{tabular}  & \begin{tabular}[c]{@{}c@{}}0.9272\\(0.015)\end{tabular} & \begin{tabular}[c]{@{}c@{}}0.9917\\(0.002)\end{tabular} & \begin{tabular}[c]{@{}c@{}}0.9585\\(0.007)\end{tabular} & \begin{tabular}[c]{@{}c@{}}0.9282\\(0.014)\end{tabular} & \begin{tabular}[c]{@{}c@{}}0.9918\\(0.001)\end{tabular}  \\ 
\hline
M4                        & \begin{tabular}[c]{@{}c@{}}0.9578\\(0.006)\end{tabular}  & \begin{tabular}[c]{@{}c@{}}0.9333\\(0.011)\end{tabular} & \begin{tabular}[c]{@{}c@{}}0.9919\\(0.001)\end{tabular} & \begin{tabular}[c]{@{}c@{}}0.9587\\(0.006)\end{tabular} & \begin{tabular}[c]{@{}c@{}}0.9339\\(0.010)\end{tabular} & \begin{tabular}[c]{@{}c@{}}0.9918\\(0.001)\end{tabular}  \\ 
\hline
F1                        & \begin{tabular}[c]{@{}c@{}}0.9579\\(0.012)\end{tabular}  & \begin{tabular}[c]{@{}c@{}}0.8914\\(0.045)\end{tabular} & \begin{tabular}[c]{@{}c@{}}0.9918\\(0.002)\end{tabular} & \begin{tabular}[c]{@{}c@{}}0.9583\\(0.011)\end{tabular} & \begin{tabular}[c]{@{}c@{}}0.8944\\(0.042)\end{tabular} & \begin{tabular}[c]{@{}c@{}}0.9917\\(0.001)\end{tabular}  \\ 
\hline
F2                        & \begin{tabular}[c]{@{}c@{}}0.9606\\(0.009)\end{tabular}  & \begin{tabular}[c]{@{}c@{}}0.9293\\(0.016)\end{tabular} & \begin{tabular}[c]{@{}c@{}}0.9939\\(0.001)\end{tabular} & \begin{tabular}[c]{@{}c@{}}0.9596\\(0.008)\end{tabular} & \begin{tabular}[c]{@{}c@{}}0.9280\\(0.16)\end{tabular}  & \begin{tabular}[c]{@{}c@{}}0.9940\\(0.001)\end{tabular}  \\ 
\hline
F3                        & \begin{tabular}[c]{@{}c@{}}0.9557\\(0.011)\end{tabular}  & \begin{tabular}[c]{@{}c@{}}0.9088\\(0.019)\end{tabular} & \begin{tabular}[c]{@{}c@{}}0.9946\\(0.004)\end{tabular} & \begin{tabular}[c]{@{}c@{}}0.9545\\(0.011)\end{tabular} & \begin{tabular}[c]{@{}c@{}}0.9084\\(0.017)\end{tabular} & \begin{tabular}[c]{@{}c@{}}0.9946\\(0.003)\end{tabular}  \\ 
\hline
F4                        & \begin{tabular}[c]{@{}c@{}}0.9643\\(0.008)~\end{tabular} & \begin{tabular}[c]{@{}c@{}}0.9062\\(0.014)\end{tabular} & \begin{tabular}[c]{@{}c@{}}0.9959\\(0.002)\end{tabular} & \begin{tabular}[c]{@{}c@{}}0.9644\\(0.007)\end{tabular} & \begin{tabular}[c]{@{}c@{}}0.9063\\(0.013)\end{tabular} & \begin{tabular}[c]{@{}c@{}}0.9958\\(0.001)\end{tabular}  \\
\hline
\end{tabular}
\end{table}

\section{Conclusions}
In this work, we view rtMRI synthesis from phonemes as a sequence-to-sequence problem. We use a transformer encoder and convolutional decoder \emph{s2s} to predict full sequence rtMRI video, on 8 subjects. We further condition the \emph{s2s} model with latent features from phoneme-conditioned variational autoencoder(CVAE) in order to make the training robust. We evaluate with subjective and objective metrics based on air tissue boundary(ATB) segmentation. With the least MOS score of 3.396 and least dice score of 0.89, we show that the synthesised samples are of reasonably good quality and also show the benefit of having \emph{s2s-v} modification. In the future, we plan to focus more on the objective metric, to be a more precise score, and also capture the errors in temporal continuity of different articulator contours from the synthesised videos.
\bibliographystyle{IEEEbib}
\bibliography{refs}

\end{document}